\documentclass[preprint2]{aastex6}
\usepackage{amsmath}
\usepackage{float}
\usepackage[section]{placeins}
\newcommand{\kms}{\ensuremath{{\rm km}\,{\rm s}^{-1}}}

\newcommand{\teff}{T$_{\rm eff}$}
\newcommand{\logg}{$\log{g}$}

\slugcomment{Accepted by The Astrophysical Journal on August 22, 2017}

\begin{document}

\title{HD 49798: Its History of Binary Interaction and Future Evolution}

\author{Jared Brooks\altaffilmark{1}, Thomas Kupfer\altaffilmark{2}, Lars Bildsten\altaffilmark{1,3}}

\altaffiltext{1}{Department of Physics, University of California, Santa Barbara, CA 93106}
\altaffiltext{2}{Division of Physics, Mathematics, and Astronomy, California Institute of Technology, Pasadena, CA 91125, USA}
\altaffiltext{3}{Kavli Institute for Theoretical Physics, University of California, Santa Barbara, CA 93106}

\begin{abstract}

The bright subdwarf-O star (sdO) HD 49798 is in a 1.55 day orbit with a compact companion that is spinning at 13.2 seconds.
Using the measurements of the effective temperature ($T_{\rm eff}$), surface gravity ($\log g$), and surface abundances of the sdO, we construct models to study the evolution of this binary system using Modules for Experiments in Stellar Astrophysics ($\texttt{MESA}$).
Previous studies of the compact companion have disagreed on whether it is a white dwarf (WD) or a neutron star (NS).
From the published measurements of the companion's spin and spin-up rate, we agree with Mereghetti and collaborators (2016) that a NS companion is more likely.
However, since there remains the possibility of a WD companion, we use our constructed $\texttt{MESA}$ models to run simulations with both WD and NS companions that help us constrain the past and future evolution of this system.
If it presently contains a NS, the immediate mass transfer evolution upon Roche lobe (RL) filling will lead to mass transfer rates comparable to that implied in ultraluminous X-ray sources (ULXs).
Depending on the rate of angular momentum extraction via a wind, the fate of this system is either a wide ($P_{\rm orb}{\approx} 3$ day) intermediate mass binary pulsar (IMPB) with a relatively rapidly spinning NS (${\approx} 0.3$ s) and a high mass WD (${\approx} 0.9 M_\odot$), or a solitary millisecond pulsar (MSP).

\end{abstract}


\section{Introduction}

HD 49798 is a binary system consisting of a hot and bright subdwarf in a $1.55$ day orbit with a compact companion.
At the time of its discovery and classification it was the brightest hot subdwarf known \citep{Jaschek1963}, and remains one of the brightest today \citep{Mereghetti2011}.
It was initially known to be a binary system, but \cite{Thackeray1970} was the first to give a spectroscopic orbital period of $1.5477$ days and to suggest that the compact companion may be a WD.
Just two years later \cite{Dufton1972} performed a non-local thermodynamical equilibrium (LTE) analysis to derive estimates on the effective temperature ($T_{\rm eff}$) and the surface gravity ($\log g$) of the sdO star, and \cite{Kudritzki1978} improved upon these measurements; based on non-LTE modeling they found \teff=47\,500$\pm$2000\,K, \logg=4.25$\pm$0.2, $y=50^{+10}_{-7}\,\%$, where $y=n({\rm He})/n$, as well as a projected rotational velocity $v_{\rm rot}\sin$ i=45$\pm$5\.\kms for the hot subdwarf. 
This result was confirmed by an independent non-LTE analysis based on high-resolution Very Large Telescope (VLT)/Ultraviolet- Visual Echelle Spectrograph (UVES) spectra \citep{Mueller2009}. 
He found \teff=46\,500$\pm$500\,K, \logg=4.35$\pm$0.1, $y=50\,\%$ as well as a projected rotational velocity $v_{\rm rot}\sin i=42\pm$5\.\kms which agrees with \cite{Kudritzki1978}.
Additionally, \cite{Bisscheroux1997} also did analysis on the subdwarf and via a common envelope (CE) ejection efficiency parameterization concluded that an intermediate mass star that entered into a CE while on the early-AGB (EAGB) is the most likely progenitor to HD 49798.

This system was also detected in X-rays. \cite{Israel1995, Israel1996} published a detection of a $13.2$ second period X-ray pulse which is interpreted as the spin period of a magnetic compact companion accreting from the subdwarf wind.
The estimates of the mass loss rates from the subdwarf and the capture rate onto the companion and the associated accretion luminosity were compared to the observed X-ray luminosity by \cite{Israel1996} and led to their suggestion that a NS was more likely than a WD.

The layout of this paper is as follows.
We continue a short review of previous studies of this binary system and confirm that our sdO stellar model matches the observations in \S \ref{sec:obs}.
Then in \S \ref{sec:ns} we give our arguments for a NS companion and show results of binary modeling to give predictions on the future of the binary system.
We also show results for binary modeling assuming a WD companion in \S \ref{sec:wd}.
We explore the outcomes of a merger caused by a high rate of angular momentum loss via the system wind in \S \ref{sec:merger}, and finish with our conclusions in \S \ref{sec:conclusions}.

\section{Observational Analysis}\label{sec:obs}

In this study, we build stellar models that match the measured values of $T_{\rm eff}$, $\log g$, mass, and surface abundances of the sdO star and constrain the past and future evolution of this system using $\texttt{MESA}$ version 8118 \citep{Paxton2011,Paxton2013,Paxton2015a}.

\subsection{Previous Compact Object Interpretation}

\cite{Bisscheroux1997} looked at the same X-ray data from ROSAT as \cite{Israel1996}, but used different estimates for the wind mass loss rates from the sdO star and concluded that a WD is more likely, but, a NS cannot be ruled out.
One of their arguments against a NS companion has to do with their low birthrate and the small likelihood of seeing such a system.
This argument does not hold because in the alternative scenario, a WD companion would accrete enough to reach $M_{\rm Ch}$ and undergo an accretion-induced collapse (AIC), leaving a subdwarf and a NS.

Several papers from the same group have been published on this system in the past few years \citep{Mereghetti2010, Mereghetti2011, Mereghetti2013, Mereghetti2016}. 
\cite{Mereghetti2010} detected an eclipse in the X-ray light curve with a period coincident with the spectroscopic period. This allowed them to derive the inclination of the system and a much more precise measurement of the masses in the system and found $M_{\rm sdO}=1.50\pm0.05$, $M_{\rm CC}=1.28\pm0.05$, where $M_{\rm CC}$ is the mass of the compact companion.
They also use the eclipse duration to measure the size of the X-ray emitting region to be ${\approx}10^{4}$ km, which is more that two orders of magnitude larger than the blackbody radius they derive from the X-ray spectrum.
Just as in previous studies, the authors used wind-capture accretion rates and compared to the X-ray luminosity to help distinguish between a NS or WD companion.
\cite{Mereghetti2010, Mereghetti2011, Mereghetti2013} all favor a WD over a NS companion, but the new angular momentum and magnetic field analysis in \cite{Mereghetti2016} suggests that a NS companion is more likely.
\cite{Wang2010} and \cite{Liu2015} performed calculations for this system assuming a C/O WD companion and concluded that it may be a SN Ia progenitor. We did not pursue such an interpretation for two reasons: (1) C/O WDs are not expected to form above $1.05 M_\odot$ \citep{Piersanti2014}, and (2) we expect that shell carbon ignitions would transform C/O WDs to O/Ne before reaching $M_{\rm Ch}$ (see \S \ref{sec:wd}).

\subsection{sdO Modeling}

Using \texttt{MESA}, we took the $1.5 M_\odot$ measurement of the mass of the subdwarf derived from combination of the X-ray mass function and the optical mass function by \cite{Mereghetti2010} and constructed a model by starting with a $7.15 M_\odot$ zero-age main sequence (ZAMS) model, and started mass loss on the EAGB, just before the second dredge-up, and ended mass loss when the surface helium mass fraction of the model matched the observed value.

\begin{figure}[H]
  \centering
  \includegraphics[width = \columnwidth]{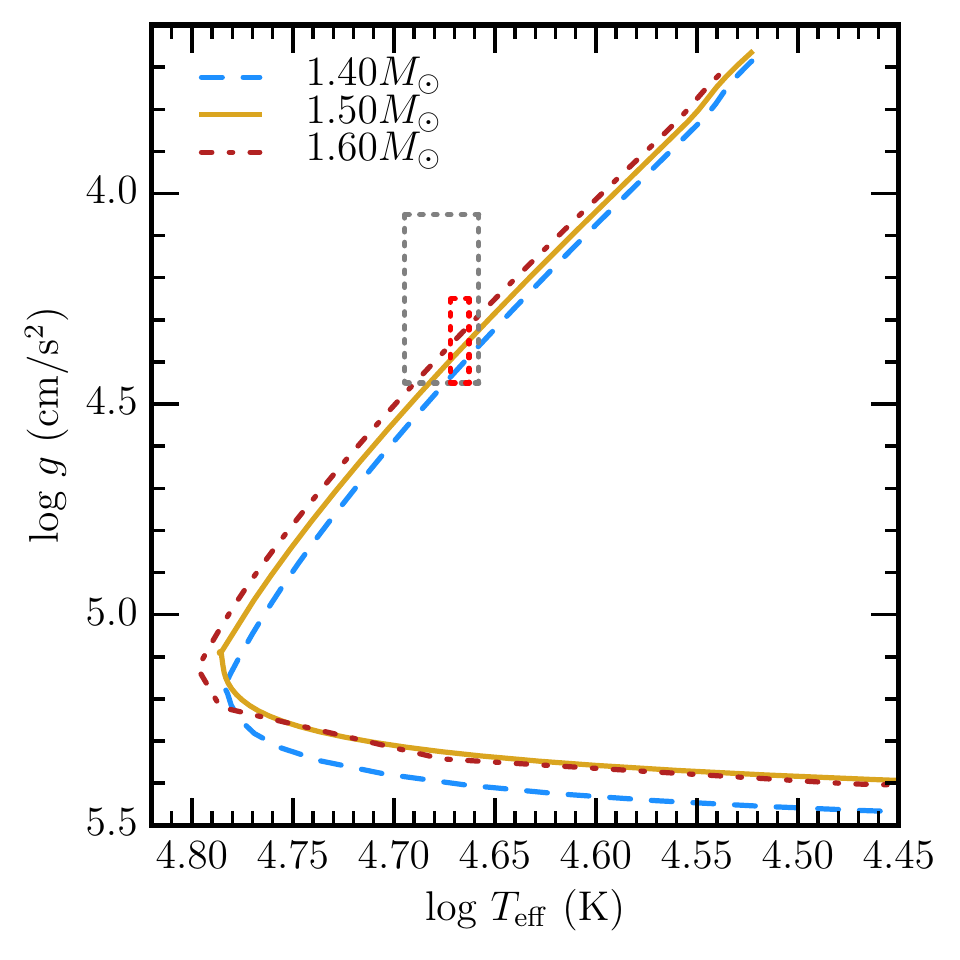}
  \caption{\footnotesize 
  The dotted gray square represents the error box of the measurements of the $T_{\rm eff}$ and $\log g$ of HD 49798 from \cite{Mereghetti2010}, and the dotted red box is from \cite{Mueller2009}.
  The curves show the evolution of Helium cores of different masses evolving from the bottom of figure and move towards lower $T_{\rm eff}$ and $\log g$ and are cut off as RLOF begins in the upper right of the plot.
  }
  \label{fig:2}
\end{figure}

\begin{deluxetable}{ccc}
  \tablecaption{Comparing to observations of the sdO\label{tab:1}}
  \tablehead{\colhead{Observable} & \colhead{Observed} & \colhead{$1.50 M_\odot$ model}}
  \startdata
  $T_{\rm eff}$ (K)  & 47,500 $\pm$ 2000   & 47,500 \\ 
  $\log g$ (cm s$^{-2}$)  & 4.25 $\pm$ 0.2   & 4.41 \\
  Radius ($R_\odot$)  & 1.45 $\pm$ 0.25 &  1.25 \\
  log Lum. ($L_\odot$)   & 3.90 $\pm$ 0.15   & 3.85 \\ 
  $X_{\rm He, surf}$ & 0.78$\pm$0.07 & 0.78 
  \enddata
\end{deluxetable}

When the effective temperature ($T_{\rm eff}$) of the subdwarf model reaches the observed $47,500$ K, the $\log g$ measurement and derived radius and luminosity all agree with the model within the given error bars, as shown in Table \ref{tab:1}.
We also show this in the $T_{\rm eff}-\log g$ diagram of Figure \ref{fig:2}.

At the time when the measurements match the model, the carbon core has grown to $0.71 M_\odot$, and the surface is blowing off a wind at $6\times10^{-9} M_\odot$/yr, using the wind prescription from \cite{Blocker1995} and a scaling factor of $0.05$.
According to the model, the star will fill its RL approximately 65,000 years from now.

\section{Neutron Star companion}\label{sec:ns}

The measured spin-up rate of $\dot{P}=-2.15\times10^{-15}$ s s$^{-1}$ given in \cite{Mereghetti2016} is high for a WD, requiring a relatively large accretion rate.
At the maximum wind mass loss rate from the donor of $10^{-8} M_\odot$/yr \citep{Hamann1981, Mereghetti2011}, the companion would need to capture all the wind that crosses its RL to cause the measured spin up \citep[see eq. 14 in][]{Mereghetti2016}, which is likely an overestimate of the wind-capture rate.
Therefore, we consider a NS interpretation of the compact companion to be much more likely. 

If the companion is indeed a NS, there are two ways it could have formed: (1) via a core collapse supernova (CCSN) from a star with a ZAMS mass of ${\gtrsim}10 M_\odot$, or (2) via AIC where the initially more massive star must have formed a O/Ne WD and subsequently accreted enough mass to reach $M_{\rm Ch}$.
For the AIC progenitor scenario, the sdO star must have been RL filling at the moment of AIC, and must be less than half-RL filling just after the AIC to match the observations.
Our calculations, based off the geometry of the system and the fact that the ejected mass from the AIC event (via neutrinos) takes with it the specific angular momentum of the WD, find that the change in the RL radius from that mass and angular momentum loss would only be about 4\%.
We ran models such that the donor was RL filling just before an AIC and is 4\% below RL filling just after the AIC, with a 1.55 day orbital period. 
Since the donor had a deeply convective envelope and was therefore responding adiabatically to mass loss, the sudden shut-off of mass loss causes the star to shrink, but only to 16\% below its RL before expanding to refill its RL, whereas we now observe it as about half RL filling.
If, however, the claimed distance of 650 pc is an underestimate, the radius of the sdO star would need to be significantly larger to match the measured $T_{\rm eff}$ and luminosity simultaneously.
The problems with this scenario is that the $T_{\rm eff}$, \logg, and luminosity are already simultaneously matched for the model discussed in \S \ref{sec:model}, implying that the radius is very near to what we show in Table \ref{tab:1} \citep{Kudritzki1978}, and this RL-filling progenitor model discussed here never matches the measured $T_{\rm eff}$ and \logg.

Additionally, although small eccentricities are expected for most post-AIC systems, this applies to systems with orbital periods in the range from 10 to 50 days \citep{Tauris2015}, whereas systems with periods of ${\sim}$a day end up in highly eccentric orbits \citep{Chen2011}.
Given the uncertainties in tidal circularization timescales for these unusual binaries, however, we cannot say whether the absence of an eccentricity provides any constraint on the origin of the compact object \citep{Strickland1994,Mereghetti2011}.

Furthermore, to achieve the measured surface H fraction in this scenario would require the fine tuning of having the AIC occur just as the last bit of the H-rich envelope was being transferred to the WD, whereas a CE removing only the H-rich envelope and leaving just a little bit of surface H is a much more likely explanation. 
Therefore, if the companion is a NS, it was most likely formed via CCSN and not AIC.

\subsection{MESA modeling of binary}\label{sec:model}

We use $\texttt{MESA}$'s binary module to evolve the model in a 1.55 day orbit with a $1.28 M_\odot$ point mass.
In our calculations, we assume that the system wind takes with it the specific angular momentum of the companion.
The 

\begin{figure}[H]
  \centering
  \includegraphics[width = \columnwidth]{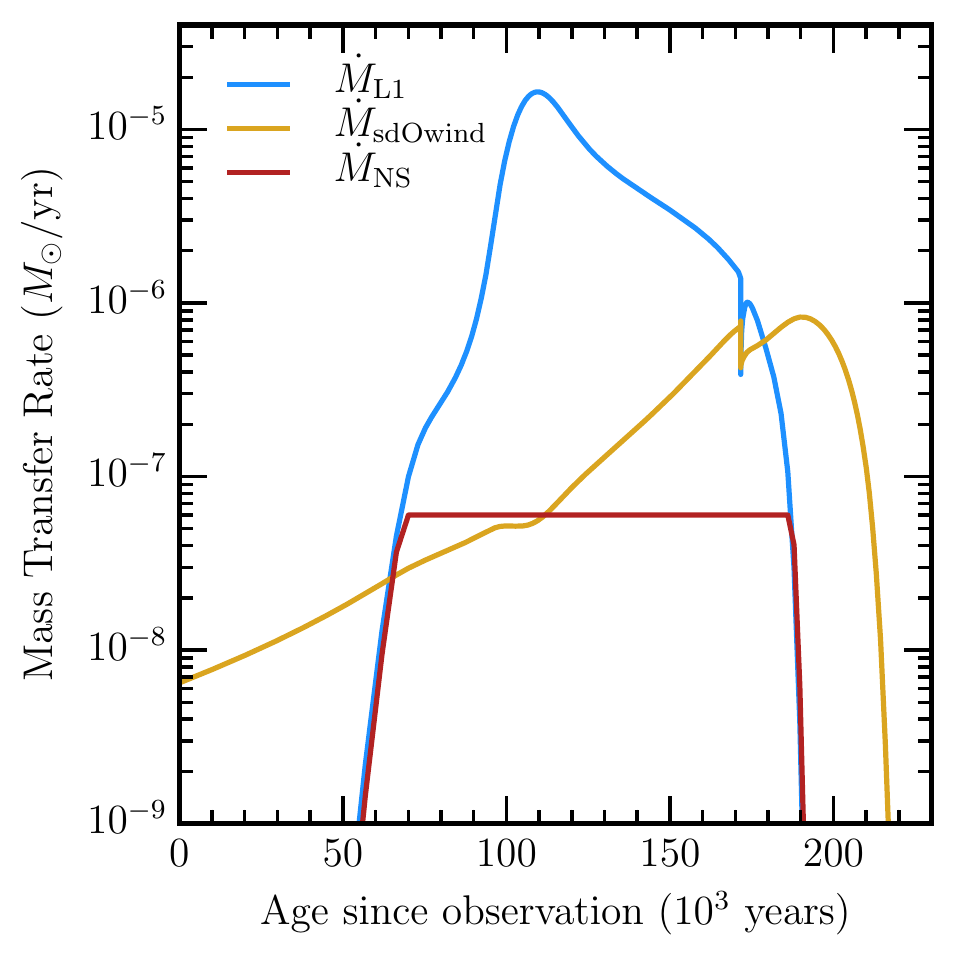}
  \caption{\footnotesize 
    The blue line shows the rate of mass transfer via RLOF.
  The orange line shows the mass loss rate from the donor via winds.
  The dark red line shows the Eddington-limited mass gain rate of the NS.}
  \label{fig:1}
\end{figure}

\noindent high system wind rates during the Roche lobe overflow (RLOF) phase will be optically thick, assuming spherical symmetry, and reprocess any high-temperature accretion luminosity.
If, however, our system-wind angular momentum assumption is a significant underestimate, then the high mass transfer rate experienced when the subdwarf begins RLOF could lead to a merger.
We briefly explore the possible outcomes of a merger between the He star and a WD or NS companion in \S \ref{sec:merger}.

At Roche lobe overflow (RLOF) the mass transfer rate through L1 quickly grows to ${\approx}2\times10^{-5} M_\odot$/yr, almost all of which is lost from the system due to the Eddington-limited accretion rate of the NS of $6\times10^{-8} M_\odot$/yr (Figure \ref{fig:1}), meaning that the NS only gains $7.4\times10^{-3} M_\odot$ during this phase, which spins up the NS to a 33 ms spin period.

\begin{figure}[H]
  \centering
  \includegraphics[width = \columnwidth]{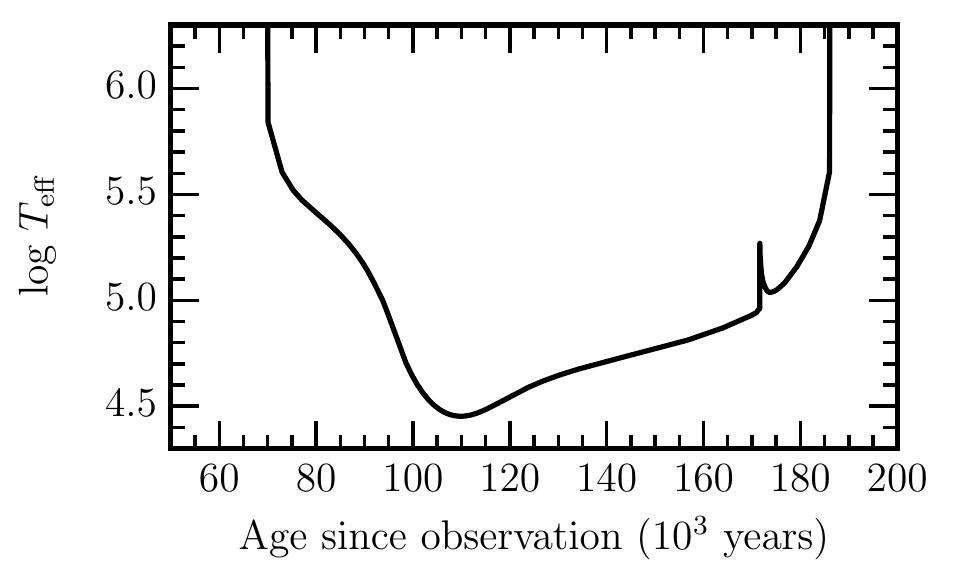}
  \caption{\footnotesize 
    Effective temperature of the optically thick wind from the NS, assuming spherical symmetry, that reprocesses the thermal X-ray radiation to lower temperatures.
    We assume electron scattering opacity, a wind speed of the escape velocity of the NS's RL, and Eddington luminosity from the NS.}
  \label{fig:4}
\end{figure}

The system loses mass at such a high rate during RLOF that the wind becomes optically thick for ${\approx}10^4$ yrs.
We compute the radius at which the optical depth of the wind reaches unity assuming a spherically-symmetric wind, a wind speed of the escape velocity of the companion's RL, and electron scattering opacity, then use that radius and Eddington luminosity of the NS to compute the $T_{\rm eff}$ that will be observed during the RLOF stage.
As shown in Figure \ref{fig:4}, the observed $T_{\rm eff}$ from the wind decreases to ${\approx}2.6\times10^4$ K at the highest mass loss rate.

When mass transfer completes and the donor star radius drops below the RL, the mass of the donor is $0.91 M_\odot$ and the orbital period is 2.7 days.
At that orbital period, the inspiral time is much longer than the Hubble time, so the fate of this system is to be a IMBP with the $0.91 M_\odot$ C/O WD made from the He star.

\subsection{Possible ULX Source}

Ultraluminous X-rays (ULX) sources are powered by accreting NSs or stellar mass black holes at rates
of ${\approx}10^{-6} M_\odot$/yr \citep{King2017}. Recent detections of persistent pulsations \citep[see ][ for a summary]{Walton2017} from many of these systems have proven that they often harbor a NS, implying that the accretion rate is ${\gtrsim}10^2$ higher than the Eddington accretion rate.
The binary evolution just described is a remarkable match for these ULX systems, as it stably provides accretion rates well above $10^{-6} M_\odot$/yr for a non-negligible amount of time (${\approx}80,000$ years) with an orbital period of 1.5 days.
To explore the outcome of this type of ULX system, we ran a NS case where we set the maximum accretion rate of the NS to $\dot{M}_{\rm max}=10^{-6} M_\odot$/yr.
This does not qualitatively change the fate of the system (final donor star mass and orbital period after end of mass transfer are roughly the same) but the NS would reach an even more rapid rotation rate, up to 3 ms (rather than 33 ms), due to the additional accreted material.
This is a reasonable scenario for this accretion rate, as the estimated  maximum magnetic field strength of $B_s\lesssim8.9\times10^{9}$ G \citep{Mereghetti2016} would not allow for a magnetosphere to form outside the NS.

\section{WD companion}\label{sec:wd}

\begin{figure}[H]
  \centering
  \includegraphics[width = \columnwidth]{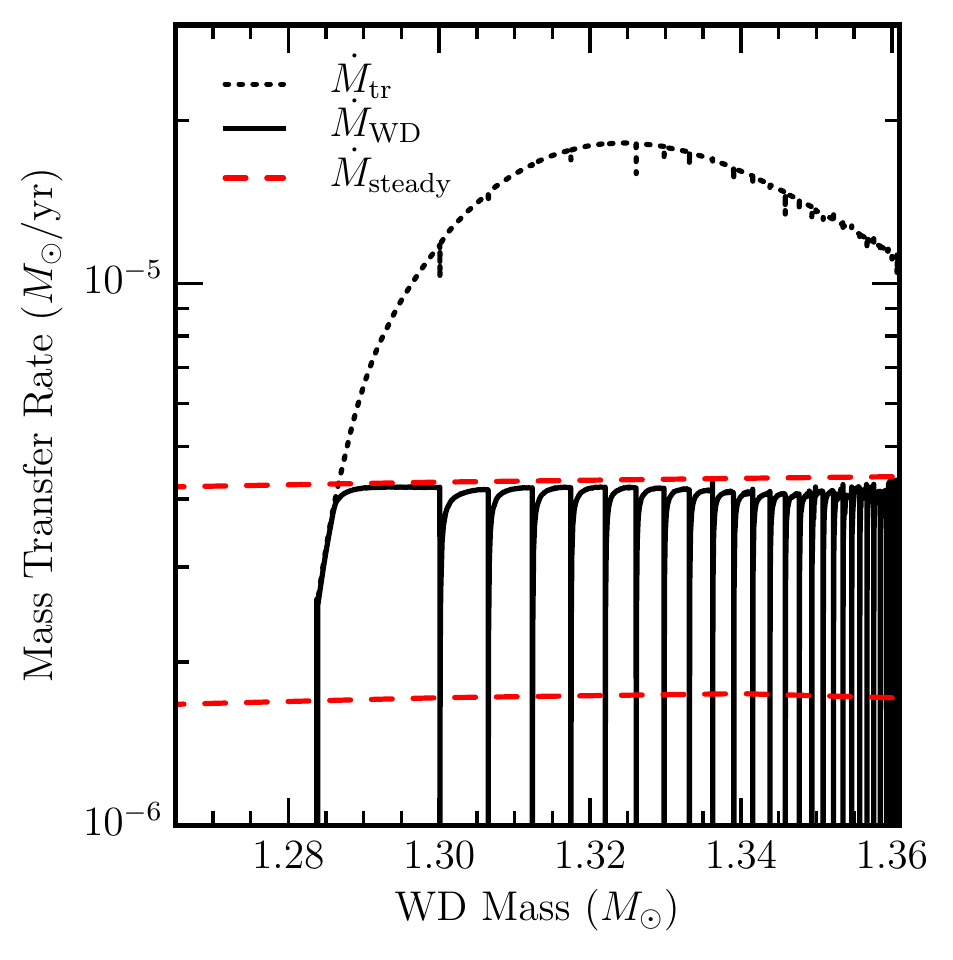}
  \caption{\footnotesize 
  Mass transfer rate of a $1.28 M_\odot$ O/Ne WD in a 1.55 day orbital period binary system with a $1.5 M_\odot$ He star shown in solid black, which is punctuated by brief mass loss episodes caused by carbon flashes in the helium burning ashes.
  The solid tracks are the rate at which the WD is gaining mass; the dotted tracks are the rate at which the He star is losing mass. 
  The difference between the dotted and solid track represents the mass that is lost from the system. 
  The stable helium burning boundaries are shown by the dashed red lines from \cite{Brooks2016}.}
  \label{fig:3}
\end{figure}

If we model the compact companion as a massive WD instead of a NS, the evolution of the He star is identical up until the start of RLOF, and extremely similar afterwards due to the mostly negligible difference in mass retention rates between the given scenarios.
The maximum accretion rate of WDs is a few orders of magnitude higher than that for a NS, so a larger fraction of the mass donated by the He star remains in the system, but still more than half the mass is ejected (see Figure \ref{fig:3}), taking with it the specific orbital angular momentum of the WD, as in \cite{Brooks2016}.
The WD steadily burns He on the surface to C/O, building up hot C/O layers that become unstable to runaway burning.
The majority of energy released from carbon burning goes into neutrinos (which free stream out of the system) and increasing the entropy of the material, lifting the degeneracy of the C/O layer and expanding the surface of the WD by a factor of $\approx2$. This mild expansion is not enough to power mass loss, but is enough to temporarily prevent surface accretion.
This can be seen in Figure \ref{fig:3}, which shows the mass transfer rate by the dotted black line and the WD mass accretion rate by the solid black line, which has sharp dips caused by the carbon burning flashes \citep{Brooks2017}.

The WD grows in mass up to $1.36 M_\odot$ when electron captures at the center begin to remove pressure support, leading to a collapse of the WD into a NS \citep{Nomoto1987, Nomoto1991, Woosley1992, Dessart2006, Schwab2015}.
At this stage, the He star has decreased in mass to $1.206 M_\odot$ and the period has increased to 1.9 days.
The orbital period increases because, although the ejected mass carries away angular momentum, the conservation of angular momentum of the transferred mass from the donor to the accretor has the net effect of increasing the orbital period.
The WD loses a significant amount of mass to neutrinos during the collapse to a NS, causing a sudden increase in the RL of the He star.
The He star quickly refills its RL due to He shell burning, and the system closely resembles that of the start of the NS case (\S \ref{sec:ns}),  with a smaller He star mass and a longer orbital period.
When the mass transfer completes and the newly formed WD drops below the RL, the mass of the WD is $0.91 M_\odot$, same as the NS scenario is \S \ref{sec:ns}, and the NS has gained $4.5\times10^{-3} M_\odot$.
The final orbital period after mass transfer completes is about 2.7 days, same as the NS scenario; the inspiral time is much longer than the Hubble time, leaving this system as an IMBP.

Just as in the NS scenario in \S \ref{sec:ns}, the donor transfers mass much faster than the companion can accept it, so the system wind is optically thick and reprocesses the thermal X-ray radiation from the WD at $T_{\rm eff}\gtrsim 10^6$ K to about $4\times10^4$ K at peak mass transfer rates.

\section{Merger Scenarios at RLOF}\label{sec:merger}

For the simulations shown in \S \ref{sec:model}, the system mass loss reaches $2\times10^{-5} M_\odot$/yr.
Since we assume that the system wind takes with it the specific angular momentum of the compact companion, the mass transfer is stable and leads to an overall increase in the orbital period and the binary separation.
If, however, this is an underestimate of specific angular momentum of the system wind, the large system mass loss rates and, thus, the large angular momentum loss rates could lead to a merger.

As mass transfer rates rise due to helium shell burning in the donor causing the star to rapidly expand into its RL, a large angular momentum loss rate will cause the size of the RL to shrink, increasing the mass loss rate, causing even faster angular momentum loss, resulting in runaway mass transfer.
The helium from the donor would form a CE, leading to a merger between the compact companion and the $0.71 M_\odot$ C/O core of the donor.

If the compact companion is a NS, we can use the results from \cite{Metzger2012} to predict the general outcome of such a merger.
We use his model NS\_C-O\_1, as its parameters are similar to HD 49798.
Using the wind rate and velocity from the disk and the total mass of the disrupted C/O core ($0.71 M_\odot$), we can estimate the total kinetic energy deposited by this disk wind.
Comparing this to the binding energy of the helium, now in a CE, we find that the energy from the disk wind is certainly large enough to eject all of the helium from the system.
Most of the disk mass from the disrupted C/O core is blown off in the disk wind, and only ${\approx}0.11 M_\odot$ is deposited on the NS, according to estimates from the NS\_C-O\_1 model from \cite{Metzger2012}.
This results in a NS of mass $M\approx 1.39 M_\odot$ with a spin period of $P_{\rm spin}\approx 2.4$ ms, making this a millisecond pulsar of average NS mass.

If the compact companion is a WD, there would be a similar lead up to the disruption of the C/O core of the donor, but the remnant would last much longer.
The disrupted C/O sitting on top of the $1.28 M_\odot$ O/Ne core would go through a viscous phase and a carbon burning flame as outlined in \cite{Schwab2016}, but since the WD would be primarily O/Ne, the flame would quench once it reached the O/Ne core, preventing the lifting of degeneracy of the core.
After the C/O burns to O/Ne and becomes part of the degenerate core, the core mass grows above $M_{\rm Ch}$ and electron captures start to relieve pressure in the center (before reaching conditions for neon burning, \citealp[see ][]{Schwab2016}), leading to an AIC.

\section{Conclusions}\label{sec:conclusions}

We have shown that the observations of the sdO star HD 49798 are well fit by a star born with a $7.15 M_\odot$ ZAMS mass that enters into a CE just before the second dredge-up on the early AGB to become a $1.50 M_\odot$ He star with a $10^{-2} M_\odot$ H-rich envelope.
Furthermore, the observations of the compact companion's X-ray pulsations suggest that a NS interpretation is more likely, but that a WD interpretation cannot be ruled out.
We used $\texttt{MESA}$ to simulate the evolution of this system, and predict that, for either a NS or WD companion, the fate of this system is to become a wide IMBP with a high mass (${\approx}0.9 M_\odot$) WD and relatively rapidly spinning NS.
This result assumes that the system wind takes with it the specific angular momentum of the compact companion.
If, however, the system wind extracts extra angular momentum, the high mass transfer rates ($2\times10^{-5} M_\odot$/yr) can lead to mergers (see \S \ref{sec:merger}).
In the event of a merger, whether the companion is a WD or NS, the predicted fate is to become a solitary NS, which would be a MSP in the case of a NS companion.
In the case of a WD companion going into a merger, the resulting spin period after an AIC inside of an extended helium envelope is uncertain.

If the companion is a NS we have shown that, during accretion, the system may have properties consistent with an ULX. So far only two ULX systems have a confirmed companion. P13 has a confirmed blue supergiant donor of spectral type B9Ia \citep{Motch2014}. The accretor in P13 shows a ${\approx}0.42$\,s slowly spinning up period which demonstrates that the accretor in P13 is a neutron star \citep{Fuerst2016,Israel2017}. The second known system, M101 ULX-1, has a Wolf-Rayet star donor in an $8.2$\,day orbit. 
The accretor is most likely a stellar mass black hole \citep{Liu2013}. 
Because of the lack of confirmed donor stars in ULX an sdO donor cannot be excluded in other ULXs. 
Therefore, we conclude that HD\,49798 is a plausible progenitor binary to a ULX. 

We thank Jim Fuller, Sterl Phinney, and Josiah Schwab for helpful conversations.
This research is funded  by the Gordon and Betty Moore Foundation through Grant GBMF5076 and was also supported by the National Science Foundation under grant PHY 11-25915.
We acknowledge stimulating workshops at Sky House where these ideas germinated.

\bibliographystyle{apj}
\bibliography{sdo_refs}

\end{document}